\documentclass[10pt,prd,superscriptaddress,amsfonts,amssymb,amsmath,showpacs,twocolumn,nofootinbib, aps]{revtex4-1}
\usepackage{bm}
\usepackage{amsfonts}
\usepackage{latexsym}
\usepackage[latin1]{inputenc}
\usepackage{graphicx}
\usepackage{amsmath}
\usepackage{palatino}
\usepackage{mathpazo}
\usepackage{textcomp}
\usepackage{float}
\usepackage{booktabs}
\usepackage{dcolumn}
\usepackage{multirow}
\usepackage{ragged2e}
\usepackage{hyperref}
\hypersetup{colorlinks,citecolor=blue}
\usepackage{amsmath}
\usepackage{xcolor}
\usepackage{orcidlink}
\usepackage[caption=false]{subfig}
\usepackage{commath}
\usepackage{comment}
\usepackage[normalem]{ulem}
\allowdisplaybreaks[1]

\allowdisplaybreaks

\newcommand{\be}{\begin{equation}}
\newcommand{\ee}{\end{equation}}
\newcommand{\bea}{\begin{eqnarray}}
\newcommand{\eea}{\end{eqnarray}}

\begin{document}

\title{Singularity resolution in spherically reduced 2D semiclassical gravity\\ with negative central charge}

\author{Adri\'an del R\'io}
\email{adrdelri@math.uc3m.es}
\affiliation{Departamento de Matem\'aticas, Universidad Carlos III de Madrid,  Avda de la Universidad 30, 28911 Legan\'es, Spain.}
\affiliation{Departamento de F\'isica Te\'orica and IFIC, Centro Mixto Universidad de Valencia-CSIC. Facultad de F\'isica, \\ Burjassot-46100, Valencia, Spain.}
\author{F. Javier Mara\~{n}\'on-Gonz\'alez}
\email{jmarag@ific.uv.es}
\affiliation{Departamento de F\'isica Te\'orica and IFIC, Centro Mixto Universidad de Valencia-CSIC. Facultad de F\'isica, \\ Burjassot-46100, Valencia, Spain.}
\author{Jos\'e Navarro-Salas}
\email{jnavarro@ific.uv.es}
\affiliation{Departamento de F\'isica Te\'orica and IFIC, Centro Mixto Universidad de Valencia-CSIC. Facultad de F\'isica, \\ Burjassot-46100, Valencia, Spain.}

\date{\today}
\begin{abstract}
We analyze the semiclassical Schwarzschild geometry in the Boulware quantum state in the framework of two-dimensional (2D) dilaton gravity. The classical model is defined by the spherical reduction of Einstein's gravity sourced with conformal scalar fields.  The expectation value of the stress-energy tensor in the Boulware state  is singular at the classical horizon of the Schwarzschild spacetime, but when  backreaction effects are considered, previous results have shown that the 2D geometry is horizonless and described by a nonsymmetric wormhole with a curvature singularity on the other side of the throat. In this work we show that reversing the sign of the central charge of the conformal matter removes the curvature singularity of the 2D backreacted geometry, which happens to be  horizonless and asymptotically flat. This result is consistent with a similar analysis recently performed for the Callan-Giddings-Harvey-Strominger model. We also argue the physical significance of negative  central charges in  conformal anomalies from a four-dimensional perspective.


\end{abstract}

\maketitle

\section{Introduction}

Two-dimensional (2D) dilaton gravity has proven to be a  useful theoretical framework for exploring quantum effects in gravity and black hole physics. By reducing the complexity inherent in general relativity and in higher-dimensional theories, 2D  models provide a more tractable arena for studying fundamental aspects of black hole formation and evaporation, including the role of horizons, singularities, one-loop semiclassical effects and the quantum nature of gravity.

Interest in 2D dilaton gravity was initially stimulated by the introduction of the Jackiw-Teitelboim model \cite{J, T}, and was later greatly popularized  by the Callan-Giddings-Harvey-Strominger (CGHS) model \cite{CGHS}. In addition to the spacetime metric $g_{ab}$ describing the gravitational field, these black hole models  involve a scalar field $\phi$, known as the dilaton field, whose coupling with $g_{ab}$ makes the 2D theory nontrivial, and therefore plays a crucial role in the dynamics of the theory.  In particular, the classical CGHS action is given by
\be \label{CGHSaction}
S_{CGHS}=\frac{1}{2G_2}\int d^2 x \sqrt{-g}\, e^{-2 \phi}(R+4(\nabla \phi)^2+4 \lambda^2) + S_m \ , 
\ee
where $R$ is the Ricci scalar curvature, $G_2$ plays the role of the two-dimensional Newton's  constant, $\lambda$ is a cosmological constant  term, and \bea
S_m= -\frac{1}{2}\sum_{i=1}^N\int d^2x \sqrt{-g}\, \nabla_a f_i \nabla^a f_i\, ,
\eea
is the matter action, described by some scalar  fields $\{f_i\}_{i=1}^N$. This model has been instrumental in exploring  quantum aspects of black holes and the  modeling of black hole evaporation due to Hawking radiation, including questions regarding information loss \cite{Hawkingcghs, Harvey, Strominger} (see also the review \cite{review2D} and the monograph \cite{FN05}).

The CGHS model was inspired from string theory. A similar 2D dilaton gravity model can be obtained from the dimensional reduction of four-dimensional general relativity in spacetimes of spherical symmetry. More precisely, by integrating out the angular degrees of freedom in the Einstein-Hilbert action for spherically symmetric metrics,  an effective 2D theory arises, where  the radial coordinate in the original four-dimensional spacetime plays the role of the dilaton field in the reduced theory. The resulting action can be written as \cite{FN} 
\be \label{Einsteinspherical}
S_{GR2d}=\frac{1}{2G_2}\int d^2 x \sqrt{-g}e^{-2 \phi}\left[\frac{R}{2}+(\nabla \phi)^2+ \frac{e^{2 \phi}}{r_0^2}
\right]+S_m , 
  \ee
where $\phi$  is now given by the logarithm of the  radial coordinate and $r_0$ is a length scale parameter. Notice the close similarity of this dilaton gravity theory with respect to (\ref{CGHSaction}).
Because the causal structure of spherically symmetric black holes in general relativity is two-dimensional, this dimensional reduction is expected to retain essential features of the  dynamics of actual black holes, while at the same time simplifying the problem to two spacetime dimensions.

Upon quantization of the matter fields $f_i$ on a 2D spacetime background,  it is possible to obtain an explicit formula for the renormalized vacuum expectation value of the stress-energy tensor, $\langle T_{ab}\rangle$. Roughly speaking, this is possible by integrating the two-dimensional conformal or trace anomaly $\langle T^a_a\rangle = \frac{C\, \hbar }{24\pi} R$, where $C$ is the central charge of the matter conformal field theory. It is then possible to incorporate the contributions from the quantum vacuum  in an effective action. The full effective action for the dilaton gravity model in the semiclassical regime is obtained either from (\ref{CGHSaction}) or (\ref{Einsteinspherical}) after doing the substitution: 
\be \label{semiclassical}
S_m \to -\frac{C\hbar }{96 \pi} \int d^2 x \sqrt{-g} R \, \square^{-1} R + S_{local}\, .
\ee
This is called the Polyakov action \cite{P}. 
The nonlocal term $R \Box^{-1} R$ accounts for the quantum corrections due to the 2D conformal anomaly, while $S_{local}$ represents  a generic local counterterm. The ambiguities in the nonlocal operator can be understood  as the freedom in the choice of the quantum vacuum state,  in the description of the quantum theory.

 In this work, we will focus on static and asymptotically flat solutions of the semiclassical field equations derived from (\ref{semiclassical}), for the model  (\ref{Einsteinspherical}). Notable static states linked to static semiclassical solutions are the Hartle-Hawking and the Boulware states. The first describes a black hole in  equilibrium with a thermal bath of radiation. The corresponding vacuum expectation value of the stress tensor $\langle T_{ab}\rangle$ is nonzero at infinity (it agrees with the expected thermal radiation), and  is regular at the event horizons.\footnote{The Unruh state is also regular at the future event horizon, but it is singular at the past horizon. This asymmetry models the effect of a  gravitational collapse \cite{FN, FN05}.} In contrast, $\langle T_{ab}\rangle$ in the Boulware state decays to zero at infinity, and is singular, for a fixed classical black hole, at the horizons. When including backreaction effects, this suggests significant deviations   in the vicinity of the classical horizon \cite{CF, Candelas80}.  This can be analytically confirmed in the CGHS model.{\footnote{The CGHS model has a classical global symmetry. The intrinsic ambiguities for local counterterms $S_{local}$ in the one-loop effective action can be fixed by choosing a specific local counterterm to maintain, at the quantum level, the referred symmetry. This defines the so-called RST model \cite{RST} as a specific semiclassical model associated to the classical CGHS model \cite{CN}. For a different  approach, see \cite{Ashtekar-Pretorius}.} More precisely, the semiclassical static solutions of (\ref{semiclassical}) for   (\ref{CGHSaction}) happen to be horizonless and asymptotically flat. In fact, the classical horizons are removed and converted into null curvature singularities \cite{FN05, Zas}. Interestingly, recent works \cite{PSSprl, PSS1, PSS2} have shown that the curvature singularities of the CGHS model can be  removed, provided the central charge of the matter fields is assumed negative. The resulting two-dimensional spacetime appears geodesically complete and horizonless, and it has the same causal structure of the two-dimensional Minkowski spacetime. 
The apparent  inconvenience for using a conformal field theory with a negative central charge, such as ghosts,  is circumvented by arguing that the vacuum stress tensor vanishes in the asymptotically flat regions (forced by the Boulware state condition). Conformal matter with a negative central charge, and hence nonunitary, is allowed to participate in the gravity theory, provided that it does not appear in the asymptotic regions \cite{PSSprl, PSS1, PSS2}.    Moreover, from a higher dimensional perspective  
there are additional reasons to explore the geometrical consequences of this assumption. 
Namely, in 3+1 dimensions it is possible to have conformal field theories with negative central charges that respect unitarity (see Sec. \ref{Conclusions}).

On the other hand, the analysis of the semiclassical theory in the Boulware state for the model (\ref{Einsteinspherical}) was initiated in \cite{FFNOS06}. The model is not analytically solvable and the study of the semiclassical solutions required numerical techniques. The central charge for conformal matter was assumed to be positive, and the resulting backreacted geometry  happens to be horizonless and with a null curvature singularity, just as in the CGHS model. Physically, the resulting quantum corrected Schwarzschild geometry describes an asymmetric wormhole. Namely, an asymptotically flat branch connects the throat, and a null singularity develops beyond it.  This qualitative picture agrees with  the results of more recent works \cite{Ho, Julio20}, and  with a similar  analysis carried out in 3+1 dimensions, using the four-dimensional conformal anomaly of scalar fields \cite{Pau, Julio2}. In the present work, we will show that the semiclassical, backreacted Schwarzschild geometry in the Boulware state for the model (\ref{Einsteinspherical}), and with {\it a negative central charge}, is  free of curvature singularities.  The two-dimensional spacetime is regular and asymptotically flat.  

The paper is organized as follows. In Sec. II we will briefly review the analytical results for the semiclassical CGHS model. In Sec. III we will present our analysis for the spherically reduced Einstein's theory described by (\ref{Einsteinspherical}).  In Sec. IV we discuss our results emphasizing potential interpretations from a four-dimensional perspective.

\section{Boulware state in the semiclassical CGHS model}
\label{sectioncghs}

To better understand the results of  the next section, and to put them in a broader context, it is useful to  review and summarize the results of the Boulware state in the semiclassical CGHS model. 

As mentioned in the previous section, the dynamics of the classical CGHS theory is given by the action (\ref{CGHSaction}), while the quantum corrections  are obtained by integrating out the matter fields in the effective action, yielding as a result the effective Polyakov action (\ref{semiclassical}). In general, this semiclassical theory cannot be solved in full closed form. However, it turns out that it is possible to construct an analytically solvable theory if we choose $S_{local}$ in (\ref{semiclassical}) such that
\be S_{local} = S_{RST}\equiv  \frac{-\hbar C}{48\pi}\int d^2x \sqrt{-g} \phi R \ . \ee
This is the RST version of the semiclassical CGHS model \cite{RST}. 
The full semiclassical theory is described then by the effective action
\bea
S_{CGHS}&\to& \frac{1}{2G_2}\int d^2 x \sqrt{-g}[e^{-2 \phi}(R+4(\nabla \phi)^2+4 \lambda^2) \nonumber\\
&&-  \frac{C\, \hbar}{96 \pi }\int
d^2x \sqrt{-g}(R\, \Box^{-1} R+2\phi R)\ ,
\eea
where $\Box^{-1}$ is  the inverse of the linear operator $\Box=g^{ab}\nabla_a \nabla_b$. The semiclassical field equations can be obtained by taking functional variations with respect to the metric $g_{ab}$ and dilaton field $\phi$. The result of doing this produces, respectively, the following equations: 
\bea
&&\left[2e^{-2 \phi}+\frac{C\hbar G_2}{24 \pi}\right]\left[2 \partial_{ \pm} \rho \partial_{ \pm} \phi- \partial_{ \pm}^2 \phi\right] 
\nonumber \\
&&=\frac{C \hbar G_2}{12 \pi}\left[\left(\partial_{ \pm} \rho\right)^2-\partial_{ \pm}^2 \rho+ t_{\pm}\right] \, ,
\eea
\bea
-\partial_{+} \partial_{-} e^{-2 \phi}-\lambda^2 e^{2(\rho-\phi)}=\frac{C\hbar G_2}{12 \pi} \partial_{+} \partial_{-}\left[\rho-\frac{\phi}{2}\right]\label{eqgplusminus} \ , 
\eea
\bea &&2 e^{-2 \phi} \partial_{+} \partial_{-}(\phi-\rho)-\partial_{+} \partial_{-} e^{-2 \phi}-\lambda^2 e^{2(\rho-\phi)} \nonumber \\
&&=\frac{C \hbar G_2}{24 \pi} \partial_{+} \partial_{-} \rho \  , \label{eqphi}
\eea
where the equations have been simplified in the  conformal gauge $ds^2= -e^{2\rho} dx^+dx^-$, with some null coordinates $\{x^+,x^-\}$ and conformal factor $\rho$.

 By combining (\ref{eqgplusminus}) and (\ref{eqphi}) one obtains  
 $\partial_+\partial_-(\rho -\phi) =0$, which reveals the preservation of a global symmetry even at the quantum level. 
 The general solution of this last equation reads $\rho - \phi= \frac{1}{2}(\omega_+(x^+) + \omega_{-}(x^-))$, and the specific choice of   the two chiral functions $\omega_{\pm}(x^{\pm})$ can be used to fix a particular set of conformal coordinates to work with.
On the other hand, the arbitrary chiral functions $t_{\pm}(x^{\pm})$ encapsulate the choice of the quantum state. The explicit form of $t_{\pm}(x^{\pm})$ also depends on the choice of the null coordinates, and transforms with the Schwarzian derivative rule \cite{Harvey, Strominger, review2D, FN05}.

The classical solution is obtained when $\hbar \to 0$. For Kruskal-type coordinates $x^{\pm}=X^{\pm}$,  defined by the condition $\omega_{\pm}(X_\pm)= 0$, the  solution of the field equations above can be expressed as: 
\begin{equation}
\begin{aligned}
d s^2 & =-\frac{d X^{+} d X^{-}}{\left(\frac{m}{\lambda}-\lambda^2 X^{+} X^{-}\right)}\, , \\
e^{-2 \phi} & =\frac{m}{\lambda}-\lambda^2 X^{+} X^{-} .
\end{aligned}
\end{equation}
This model mimics the causal structure of the 3+1 Schwarzschild black hole. First of all, it contains a future and a past event horizon located at $X^+=0$ and $X^-=0$, respectively. Furthermore, the  curvature scalar is $R=\frac{4m\lambda^2}{m-\lambda^3 X^+X^-}$, which reveals a curvature singularity at $\lambda^3X^+X^-=m$. The Penrose diagram of this spacetime is given in Fig. \ref{CGHSpenrose}. 

When quantum effects are included, $\hbar\neq 0$, it is convenient to solve the field equations in static and asymptotically flat null coordinates, usually denoted by $x^{\pm}=\sigma^{\pm}$, where the choice of the Boulware state simply corresponds to $t_{\pm}(\sigma^{\pm})=0$. These coordinates are defined by the condition $\omega_{\pm}(\sigma^\pm)= \sigma^{\pm}$, and are related to the Kruskal coordinates by the coordinate transformation  $\lambda X^{\pm}=\pm e^{\pm \lambda\sigma^{\pm}}$.  The semiclassical solution can still be obtained in full closed form for $\hbar \neq 0$, and is given by \cite{RST}
\bea \frac{C\hbar G_2}{24\pi}\phi +  e^{-2\phi} &=& -\lambda^2  X^+X^- - \frac{C\hbar G_2}{48\pi}\ln (-\lambda^2 X^+X^-) \nonumber \\
&&+ \frac{m}{\lambda}\, , \nonumber  \\
\rho&=&\phi \ . \eea

For a nonvanishing mass $m$, previous works have shown that  the semiclassical solution is largely deformed as compared to the classical one. For $C=1$, the event and apparent horizons\footnote{The apparent horizon is defined by the condition $\partial_+r^2\sim \partial_+ e^{-2\phi}=0$.} are destroyed,  and the null lines $X^+=0$ and $X^-=0$ are now converted into curvature singularities \cite{FN05}. The geometry has the form of a wormhole with throat at $X^+X^-= -\lambda^2/48$ \cite{PSS1}. In sharp contrast, for $C=-1$ the curvature vanishes for $X^+X^-\to 0$ \cite{PSSprl, PSS1, PSS2} and the singularity is resolved. The corresponding Penrose diagrams are illustrated in Fig. \ref{CGHSboulware} ($C>0$) and Fig. \ref{CGHSboulwarenegativa} ($C<0$).

  \begin{figure} \begin{center}\includegraphics[angle=0, width=70mm]{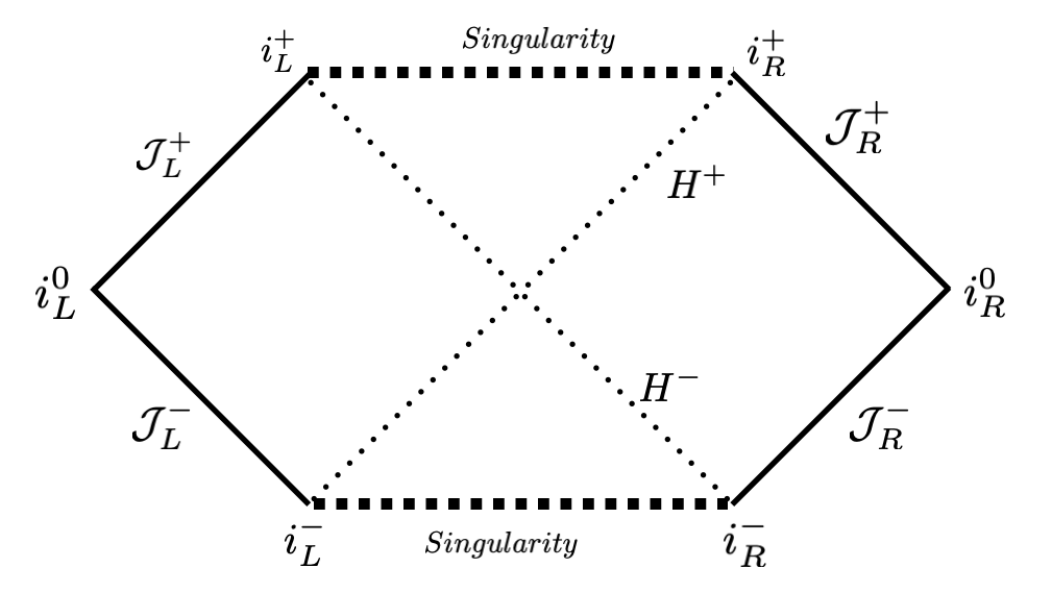}\end{center}
\caption{\small Penrose diagram showing the causal structure of the classical CGHS black hole.  The two null horizons are located at $X^\pm=0$, while the spacelike curvature singularity is  located at $\lambda^3 X^+X^-=m$, where $X^{\pm}$ are  null Kruskal coordinates.} 
\label{CGHSpenrose}\end{figure}

  \begin{figure} \begin{center}\includegraphics[angle=0, width=45mm]{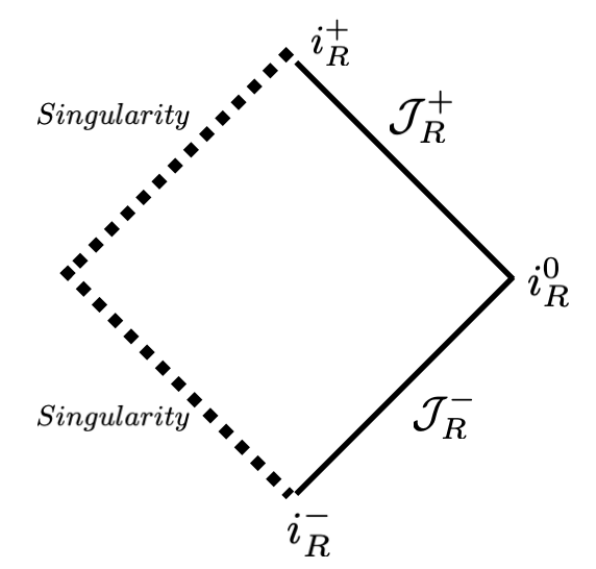}\end{center}
\caption{\small Penrose diagram showing the causal structure of the semiclassical CGHS geometry for the Boulware state and $C>0$. The two classical horizons are removed and, instead, null curvature singularities emerge  at the same locations: $X^+=0$ and $X^-=0$.} 
\label{CGHSboulware}\end{figure}

  \begin{figure} \begin{center}\includegraphics[angle=0, width=45mm]{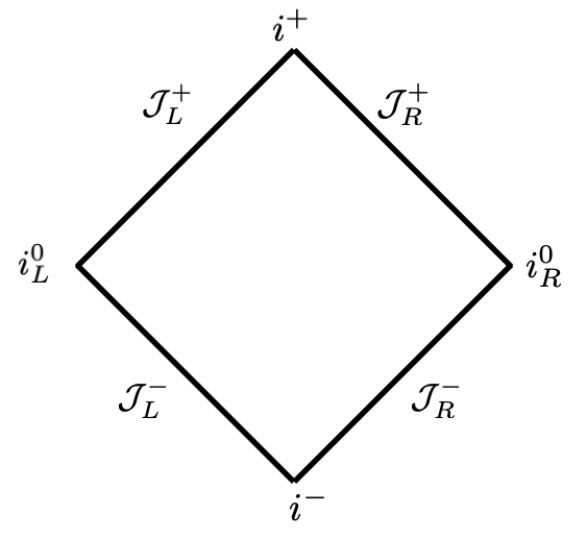}\end{center}
\caption{\small Penrose diagram showing the causal structure of the semiclassical CGHS geometry for the Boulware state and negative central charge $C<0$. The backreaction removes horizons and singularities. The global structure is similar to a two-dimensional Minkowski spacetime. $\mathcal{I}_L^-$ and  $\mathcal{I}_L^+$ are now located at the asymptotically flat regions $X^+=0$ and $X^-=0$.} 
\label{CGHSboulwarenegativa}\end{figure}

\section{Boulware state in the spherically reduced model of Einstein's gravity}
\label{sectioneinstein}

In this section we will extend the analysis carried out before for the semiclassical CGHS theory, but now using the model (\ref{Einsteinspherical}) instead. From a physical viewpoint, this is a very interesting  model because it is obtained by dimensional reduction of the standard theory of general relativity in 3+1 dimensions. 

More precisely, let us consider a spherically symmetric spacetime background in 3+1 dimensions described by general relativity. In this framework, the spacetime metric admits the following decomposition in some spherical coordinate system $(t,r,\theta,\varphi)$:
\be \label{4metric}  ds^{2}_{(4)} =  g_{ab}dx^a dx^b + r^2 d\Omega \ , \ee 
where $g_{ab}(t,r)$ is a two-dimensional metric, and $d\Omega=d\theta^2+\sin^2\theta d\varphi^2$ is the  line element on a unit-radius sphere $\mathbb S^2$. To keep the calculations simple, we will model the matter content by considering only minimally coupled, massless scalar fields $\{f_i\}_{i=1}^{N}$. The dynamics of this theory is then given by the usual Einstein-Hilbert action
\bea
S_{GR}&=&\int d^4x\sqrt{-g^{(4)}}\left[\frac{R^{(4)}}{16\pi G} - \frac{1}{2}\sum_{i=1}^N \nabla_a f_i\nabla^a f_i\right] 
\ , \ \ \ \ \ \ \ \eea
where $R^{(4)}$ is the four-dimensional Ricci scalar. 
Now, by  expanding the matter fields in spherical harmonics,  and integrating the angular variables of the action restricting to the $\ell=0$ sector, we end up with an effective two-dimensional action that reads 
\bea\label{eq:2DG}
&&\frac{1}{4G}\int d^2x\sqrt{-g}\left [r^2R+2\left(1+\nabla_a r \nabla^a r \right)\right ] \nonumber\\
&&-\frac{1}{2}\int d^2x \sqrt{-g} \, r^2\sum_{i=1}^N \nabla_a f_{i,\ell=0} \nabla^a f_{i,\ell=0} \, , \eea
where $R$ is the two-dimensional Ricci scalar, and $\nabla_a$ is the Levi-Civita connection of $g_{ab}$.
In this dimensionally reduced theory the original radial coordinate $r$ plays the role of an external scalar  field.

 The explicit coupling of $f_{i,\ell=0}$ with $r$ makes the semiclassical analysis of this theory rather intractable. To proceed further, it is customary to work in the approximation  in which the resulting two-dimensional  gravity theory is directly coupled to two-dimensional conformal matter, i.e. to keep $r^2$ constant in the second line of (\ref{eq:2DG}). In particular, with the identification $r^2=r_0^2 e^{-2\phi}$ in the first line of (\ref{eq:2DG}), we recover the action $S_{GR2d}$ in (\ref{Einsteinspherical}) with $G_2=G/r_0^2$, and $\phi$ representing the dilaton field. In order for the two-dimensional model to faithfully mimic the original four-dimensional spacetime, the field $r \propto e^{-\phi}$ is  restricted to be positive. 
 See also the recent analysis in \cite{Varadarajan}. 
 
 It is also important to note that, as opposed to  the CGHS theory, this classical action does not produce a local free field. The invariance of the classical CGHS action under the transformation $\delta \phi = \delta\rho =  e^{2\phi}$ provided the classical theory with $\partial_+\partial_-(\rho-\phi)=0$. There is no analog of this local free field for the classical spherically reduced theory.

 The semiclassical theory is constructed by replacing 
 the classical matter action in (\ref{eq:2DG}) by the corresponding semiclassical 
  Polyakov effective action: 
\bea\label{semiclassicalaction}
S_{GR2d} &\to &\frac{1}{4G}\int d^2x\sqrt{-g}\left [r^2R+2\left(1+\nabla_a r \nabla^a r\right)\right ] \nonumber\\
&&-  \frac{C\, \hbar}{96 \pi }\int
d^2x \sqrt{-g}R\, \Box^{-1} R\ , 
\eea
where $\Box^{-1}$ is again the inverse of the linear operator $\Box=g^{ab}\nabla_a \nabla_b$. The Polyakov effective action in the second line above captures the quantum aspects of the matter's vacuum state. In particular, it reproduces the vacuum expectation value of the stress-energy tensor, $\langle T_{ab}\rangle$, in the assumed approximation in which the coupling of $f_{i,\ell=0}$ with $r$ can be neglected. The ambiguity in the choice of the quantum state is encoded, precisely, in the specification of boundary data for the nonlocal operator $\Box^{-1}$. 
In contrast to the classical CGHS action, in this  new model  we have no special global conformal symmetry to preserve at the quantum level. As a consequence,  no combination of the classical field equations for $\rho$ and $\phi$ gives rise to a local free field. [One can build a very involved nonconformal symmetry, but then the corresponding counterterms in the semiclassical theory would be highly nonlocal \cite{cruzNT}]. Therefore, we have no preferred choice of local counterterm $S_{local}$ in (\ref{semiclassical}) to add. Thus, we take the simplest choice $S_{local}=0$. As we will see, due to the lack of an explicit symmetry, we end up with a theory that is significantly more difficult to solve. 

In the classical limit, $\hbar\to 0$, we recover the usual vacuum gravitational theory, a solution of which is the well-known (two-dimensional) Schwarzschild geometry, 
\bea\label{schw2d}
g_{ab}dx^adx^b=-\left[1-\frac{2 G M}{r}\right]dt^2+\left[1-\frac{2 G M}{r}\right]^{-1}dr^2\, ,
\eea 
for some $M\in\mathbb R$,  which has a curvature singularity at $r=0$: $R=\frac{4M}{r^3}$. When $\hbar\neq 0$ we can aim to solve the full field equations derived from (\ref{semiclassicalaction}) and analyze the new solution in the semiclassical regime.  Working with the dilaton field $r^2=r_0^2 e^{-2\phi}$, field variations of (\ref{semiclassicalaction}) with respect to $g^{\pm\pm}$, $g^{+-}$ and $\phi$ yield, respectively,
 \bea \label{2deinsteinsemiclassical1}
\frac{r_0^2e^{-2\phi}}{G}\left[\partial_{\pm}^2 \phi -2
\partial_{\pm} \rho
\partial_{\pm} \phi -(\partial_{\pm}\phi)^2 \right] &=& \nonumber \\ -\frac{\hbar\, C}{12\pi} \left[ (\partial_{\pm}\rho)^2
-\partial_{\pm}^2\rho + t_{\pm}\right] \ ,    \nonumber\\
 \frac{r_0^2e^{-2\phi}}{G}
\left[\partial_+ \partial_-\phi -2\partial_+\phi \partial_-\phi
 -\frac{e^{2(\rho+\phi)}}{4r_0^2} \right]
 &=& 
 \frac{\hbar\, C}{12\pi} 
\partial_+\partial_-\rho \ , \nonumber \\
 \partial_+\partial_-\rho + \partial_+\phi
 \partial_-\phi -\partial_+ \partial_-\phi&=&0 \ , 
 \  \eea
where we have expressed all the equations in conformal null coordinates $g_{ab}dx^adx^b = -e^{2\rho}dx^+dx^-$, for some scalar function $\rho$. The functions $t_{\pm}$ parametrize the ambiguity in the choice of  the quantum state. On the other hand, to simplify the equations above we will introduce the  constant $\lambda\equiv \frac{\hbar\, G}{12\pi} $, which is proportional to the square of the Planck length.

Since we are interested in exploring static solutions we will take $\rho=\rho (x)$ and $\phi=\phi(x)$, where $x= (x^+-x^-)/2$ is a spacelike coordinate. A natural quantum state in this case is the Boulware vacuum, which corresponds to $t_{\pm}(x^{\pm})=0$. We will  further assume asymptotic flatness, that is, as $x\to \infty$ the semiclassical solution should behave as the classical Schwarzschild metric:
\be ds^2 \sim \left(1-\frac{2GM}{r}\right)(-dt^2+dx^2) \, , \ee
where $M$ is an integration constant and $r^2\equiv r_0^2 e^{-2\phi}$ is related to the spatial coordinate $x$ by 
\be x\sim r + 2 GM  \log\left(\frac{r}{2GM }-1\right) \ . \ee 
In other words, asymptotically $x^{\pm}=t\pm x$ can be identified with the  Eddington-Finkelstein coordinates of the classical Schwarzschild geometry.

With all the above considerations we can rewrite the semiclassical  equations in the Boulware state as 
\begin{eqnarray}
\partial_x^2\phi_{}-(\partial_x\phi)^2-2\partial_x\rho \partial_x\phi&=&\frac{C\lambda}{r_0^2} e^{2\phi}\left(\partial_x^2\rho_{}-(\partial_x\rho)^2\right)
\label{eq:cg-i}\ , \ \ \\
\partial_x^2\phi_{}-2(\partial_x\phi)^2+\frac{e^{2(\phi+\rho )}}{r_0^2}&=& \frac{C\lambda}{r_0^2} e^{2\phi}\partial_x^2\rho_{}
\label{eq:cg-ii}\ , \\
\partial_x^2\phi_{}-(\partial_x\phi)^2-\partial_x^2\rho_{}&=&0 \label{eq:cg-iii}
 \ . \end{eqnarray} 
Note that the classical equations are recovered if $\lambda=0$. 

The above set of ordinary differential equations (ODEs) can be reduced to one single second order non-linear ODE \cite{FFNOS06}. 
In particular, it is possible to combine them and obtain 
\begin{equation}\label{eq:cg-7}
\partial_r^2\rho=-\frac{(1+r\partial_r\rho)\partial_r\rho(2r+ C\lambda  \partial_r\rho)}{r^2-
C\lambda }
 \ , \end{equation}
where the radial function $r=r_0 e^{-\phi}$ satisfies 
 \be
\label{relationrx} \left(\frac{dr}{dx}\right)^2=
\frac{e^{2\rho}}{(1+2 r \rho_r\ + C\lambda  \rho_r^2)}\ . 
\ee
At this stage, it is  useful to work with a  dimensionless radial parameter $z\equiv r/r_0$, which measures radial distance in units of $r_0$. Furthermore, it is convenient to study the inverse function $z=z(\rho )$ rather than $\rho=\rho(z)$ directly. The inverse function theorem ensures $\partial_z \rho=(\partial_\rho z)^{-1}$. A straightforward calculation shows that $z=z(\rho )$ verifies the following ODE:  \be \label{invfunc}
\partial_{\rho}^2z_{}=\frac{(z+\partial_{\rho}z)(C\frac{\lambda}{r_0^2}+2z\partial_{\rho} z)}{z^2-C\frac{\lambda}{r_0^2}}\ . \ee
In contrast to (\ref{eq:cg-7}), this is a nonlinear ODE with no explicit dependence in the independent variable, which will allow later a complementary analysis using phase space techniques. 

This ODE can now be integrated numerically on a computer.
To obtain a specific solution, we need two boundary conditions. The asymptotically flatness condition for the metric at spatial infinity imposes the following  asymptotic behavior at large $z$:
\bea\label{condin} 
\rho
(z\rightarrow\infty)\sim  \frac{1}{2}\ln\left(1-\frac{2a}{z}\right)\to 0\, , 
\eea 
where $a\equiv GM/r_0$. Inverting this functional dependence we immediately get an asymptotic condition for $z(\rho)$,
\bea \label{z1}
z(\rho\to 0)\sim \frac{2a}{1-e^{2\rho}}\to \infty\, ,
\eea
and for its derivative
\bea \label{z2}
\partial_{\rho}z(\rho\to 0)\sim \frac{4a e^{2\rho}}{(1-e^{2\rho})^2}\to \infty\, .
\eea
In practice, however, it is much more convenient to work with yet another variable, $\omega(\rho)=z^{-1}(\rho)$, since in this case the two boundary conditions for $\omega(\rho)$ at spatial infinity are finite:
\bea 
\omega(\rho=0)&=&0\, ,\label{b1}\\
\partial_{\rho}\omega(\rho=0) & = & -\frac{1}{a}= -\frac{r_0}{GM}\, .\label{b2}
\eea
This is significantly more useful for a numerical implementation. The corresponding second order ODE for $\omega(\rho)$ can be obtained by straightforward manipulation of Eq. (\ref{invfunc}): 
\bea \label{invfunc2}
\omega_{\rho\rho}=\frac{2\omega_\rho^2}{\omega}+\frac{(\omega_\rho-\omega)(C\frac{\lambda}{r_0^2}\omega^3-2\omega_\rho)}{\omega(1-C\frac{\lambda}{r_0^2}\omega^2 )}\, ,
\eea
with the notation $\omega_{\rho}\equiv \partial_{\rho}\omega$, $\omega_{\rho\rho}\equiv \partial_\rho^2 \omega$. \footnote{In the classical limit $\lambda\to 0$,  equation (\ref{invfunc2}) reduces to $\omega_{\rho\rho}=2\omega_\rho$, and it can be integrated in full closed form using boundary conditions (\ref{b1})-(\ref{b2}), leading to $\omega(\rho)=\frac{1-e^{2\rho}}{2a}$. Inverting this function to obtain $\rho=\rho(\omega)$, we recover the classical Schwarzschild geometry (\ref{schw2d}).}

As we will show in the next subsection, Eq. (\ref{invfunc2}) can be solved using  numerical techniques. With the result for $\rho=\rho(r)$ the final solution can be expressed using (\ref{relationrx}) as
\bea\label{2metric}
ds^2&=&-e^{2\rho}dt^2+(1+2r\rho_r+C\lambda \rho_r^2)dr^2\, ,\\
\phi(r)&=&-\log r/r_0\, ,
\eea
which generalizes the Schwarzschild solution in the semiclassical regime.

\subsection{Numerical solution for $C<0$}

For $C=1$ the result of  integrating (\ref{invfunc2}) with asymptotically flat conditions (\ref{b1}) and (\ref{b2}) leads to a horizonless spacetime. More precisely, the function $z(\rho)$ attains a minimum positive value, physically representing the throat of a wormhole, after  which the integration stops at a finite value of $z$, where the spacetime happens to meet a null curvature singularity \cite{FFNOS06}. The Penrose diagram is similar to Fig.  \ref{CGHSboulware}.

For $C=-1$ we need to solve the second order ODE (\ref{invfunc2}) numerically, and analyze the status of the classical horizon and curvature singularity in the new spacetime. For simplicity, we  will work in a dilaton characteristic scale  given by the Planck length, $r_0=\sqrt{\lambda}$. 

To analyze the potential curvature singularities of the solution it is necessary to work with the Ricci scalar. This geometric quantity can be expressed in terms of the radial function $z(\rho)$, or $\omega(\rho)=z^{-1}(\rho)$, as 
\be \label{2dcurvature}R =\frac{2(C\omega^3-2\omega_\rho)\omega^3}{r_0^2 (1-C \omega^2)(\omega_\rho^2-2\omega_\rho \omega+C\omega^4)}\, . \ee
This functional is zero for $\omega=0$ and any value $\omega_\rho$, so it vanishes at spatial infinity, where (\ref{b1}) and (\ref{b2}) holds. That is, our spacetime is asymptotically flat. 

From a theoretical viewpoint,  $R^{-1}(\rho_*)=0$ implies a spacetime singularity at $\rho=\rho_*$. In practical numerical implementations, however, it can be challenging to identify the emergence of curvature singularities. This is because numerical errors can hide a zero value of $R^{-1}$. If the numerical calculation produces $R^{-1}(\rho_*) \approx 0$ at some value $\rho_*$, we need a tool to distinguish  regular points (which might have high curvature, but finite) from an actual  curvature singularity that is buried below numerical uncertainties.

To analyze this issue, suppose that at some value $\rho_*$ the physical solution obtained numerically satisfies $R^{-1}(\rho_*)= r_0^2 e$, with $0 < e \ll 1$. Suppose further that the numerical calculation produces numerical errors $\epsilon_0$ and $\epsilon_1$ in $\omega$ and $\omega_\rho$, respectively. Using (\ref{2dcurvature}), a standard error analysis applied to the functional $R^{-1}= R^{-1}[\omega(\rho),\omega_\rho(\rho)]$ around the value $\rho=\rho_*$ produces
\begin{flalign}\label{num_uncertainty_r}
& \delta R^{-1}[\omega_*+\epsilon_0,\omega_{\rho_*}+\epsilon_1]= \frac{\delta R^{-1}}{\delta \omega}\epsilon_0+\frac{\delta R^{-1}}{\delta \omega_\rho}\epsilon_1+O(\epsilon_i^2)=\nonumber\\
  &r_0^2\frac{ \omega_* \omega_{\rho_*}^2(3\omega_*^2 +2 \omega_*^4-4)+(3+\omega_*^2)\omega_{\rho_*}^3 -\omega_*^7-4 \omega_*^4 \omega_{\rho_*}}{\omega_*^4(\omega_*^3+2 \omega_{\rho_*})^2}\epsilon_0\nonumber\\
  &-r_0^2\frac{(1+\omega_*^2)\omega_{\rho_*}(\omega_{\rho_*}+\omega^3_*)}{\omega_*^3(\omega^3_*+2\omega_{\rho_*})^2}\epsilon_1+O(\epsilon_0^2,\epsilon_1^2)\, .
\end{flalign}
Since $\delta R^{-1}$ tends to zero as $\epsilon_i\to 0$,  we can claim that $\rho_*$ is not a singular point in the resulting spacetime if $e$ remains nonzero in this limit.
To do the numerical computation we coded a custom numerical solver in C++ adapted to this particular problem. The code is based on a modified version of the Runge-Kutta-Fehlberg of fourth order, which uses a variable step approach.

 To make the plots more insightful and simplify the numerical analysis, we did the calculation for $a=2$, but as we will show later the qualitative behavior remains the same for astrophysically realistic values. Recall that $\lambda \equiv \hbar G/12 \pi c^3\sim 10^{-72} \text{ m}^2$, and therefore $a \equiv GM / (c^2 \sqrt{\lambda})\sim 10^9 M/$kg. In other words, our simulation corresponds to a BH of mass $\sim 10^{-8}\text{ kg}$.

\begin{figure}[htbp]
\includegraphics[angle=0,width=3.8in,clip]{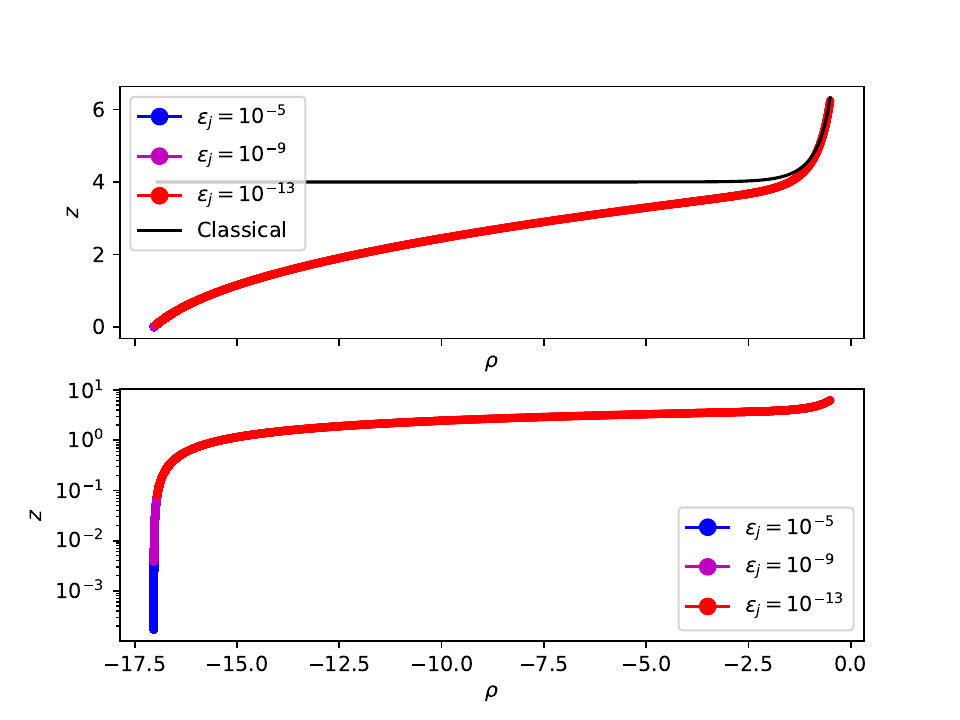}
\caption{Numerical solution for $z(\rho)=\omega^{-1}(\rho)$ obtained from the semiclassical field equations (\ref{invfunc2}) using $C=-1$, $a=2$ and asymptotically flat boundary conditions (\ref{b1}) and (\ref{b2}), with decreasing precision values $\epsilon_j$. The  classical solution (black line) shows that the function $z=z(\rho)$ 
is bounded from below,  $z>2a=4$, indicating the presence of a spacetime horizon at this location. The semiclassical solution (colored line) manages to penetrate beyond, indicating the absence of horizon in the new spacetime. The log-plot (lower panel) shows  that the solution approaches $z\to0^+$ at some finite value of $\rho$, and $z_\rho \to 1^-$ in this limit. }  
\label{fig1}
\end{figure}

\begin{figure}[htbp]
\begin{center}
\includegraphics[angle=0,width=3.8in,clip]{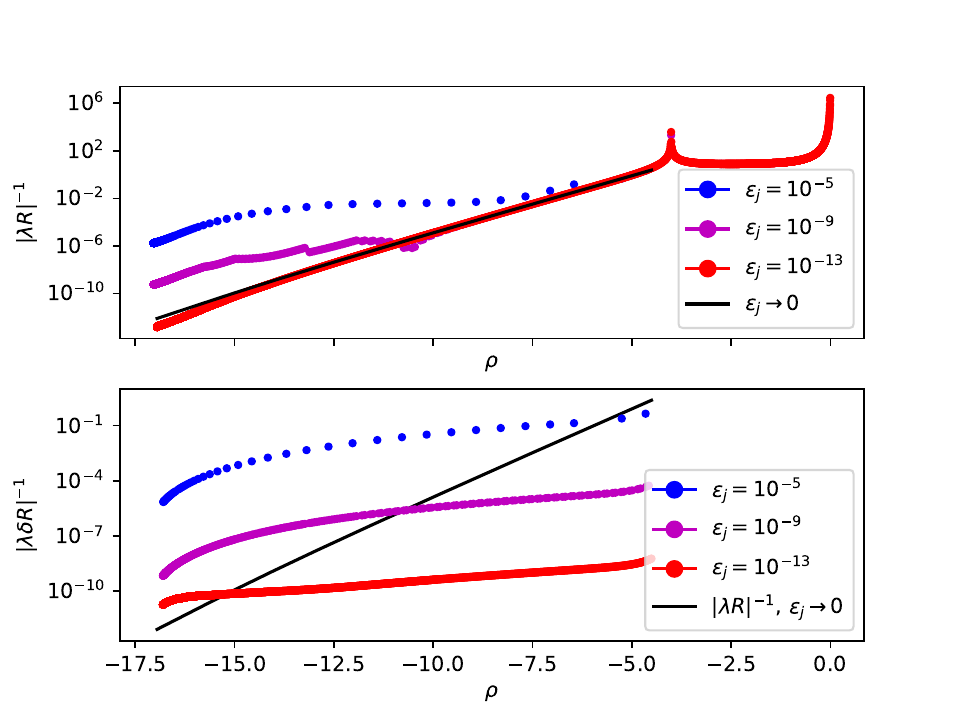}
\caption{Logarithmic plot for $|R^{-1}(\rho)|$ (upper panel) and its error (\ref{num_uncertainty_r}) (lower panel) obtained numerically from the semiclassical field equations (\ref{invfunc2}) using $C=-1$, $a=2$ and asymptotically flat boundary conditions (\ref{b1}) and (\ref{b2}), with decreasing precision values $\epsilon_j$. In the limit $\epsilon_j\to 0$ the inverse curvature converges to $R^{-1}(\rho) \to - \lambda (z^2(\rho)+1) e ^{ |m| \rho + n}/2$ around $z\sim 0$ for some real numbers $m$, $n$ (black line), while its error tends to zero. Because $\rho$ is bounded from below for all $z\geq 0$,  the curvature remains finite. That is, there is no curvature singularity in the semiclassical solution.} \label{fig2}
\end{center}
\end{figure}

We solved the ODE (\ref{invfunc2}) with asymptotically flat boundary conditions (\ref{b1}) and (\ref{b2}) for different degrees of numerical precision $\epsilon_0 = \epsilon_1$. The results are summarized in Figs. \ref{fig1} and \ref{fig2}.  Figure \ref{fig1} shows how the 
radial function $z(\rho)$ obtained approaches zero at finite $\rho$, in sharp contrast to the classical solution, whose radial function $z(\rho)$ remains always above $2a$ (the classical horizon). That is, as in the $C=1$ case, the resulting semiclassical spacetime is  horizonless. However, in contrast to the $C=1$ case, we do not find any point where the radius function  has a local minimum $(\nabla z)^2=0$, which would signal a conventional wormhole geometry, as happens for positive $C$. On the other hand,  Fig. \ref{fig2} shows that, by progressively reducing the numerical error $\epsilon_j$, the inverse curvature converges to $R^{-1}(\rho) \to - \lambda (z^2(\rho)+1) e ^{ |m| \rho + n}/2$  for some numerical constants $m,n$ (that depend of the value of $a$), while the error $\delta R^{-1}\to 0$. In other words, the classical curvature singularity at $z=0$ disappears 
when quantum effects $\lambda\neq 0$ are included.

The explicit solution for the two-dimensional metric (\ref{2metric}) is plotted in Fig. \ref{figmetric}, where the behavior of the two relevant metric components  is displayed as a function of the radial function $z$. Notice how  quantum effects regularize the coordinate singularity of the metric components at the location of the classical horizon, which is now absent. The causal structure of this solution is the same as in Minkowski space.

In 2D dilaton-gravity theories, the dilaton field $\phi$ is real valued. In the present model, constructed by dimensional reduction of spherically symmetric general relativity, this condition restricts the two-dimensional spacetime to the region $r\geq 0$. This is natural, as the role of $r$ is to represent an area radius coordinate in the 4D spacetime. Furthermore, the classical solution contained a curvature singularity at $r=0$, which prevents going further with geodesics.  However, since the semiclassical solution obtained for $C<0$ no longer displays a curvature singularity at $r=0$, it is now possible to extend analytically the spacetime domain to negative values of $r$. We will explore this in the following subsection.

\subsection{Analytical extension across $r=0$}

Since the curvature scalar does not diverge for $r=0$, the analytical extension of the metric (\ref{2metric}) through negative values of $r$ follows in a natural way by solving the ODE (\ref{invfunc}).\footnote{This analytical extension cannot be obtained by solving numerically (\ref{invfunc2}) because $z=0$ is only reached asymptotically when $\omega\to \infty$.} Such analytical extension can be obtained by evaluating the solution of (\ref{invfunc2}) obtained above at some $\omega>0$  (or $0<z<\infty$),  and using it as a boundary condition for solving (\ref{invfunc}).

The results of this integration for $z=z(\rho)$ are shown in Fig \ref{fig20}, together with the radial profile of the scalar curvature $R=R(\rho)$. Our results indicate that the scalar curvature $|R|$ has a maximum  at $r=0$ ($\rho\sim -17.5$) and then decays exponentially until the simulation stops. In other words, we report numerical evidence that when $C=-1$ the semiclassical spacetime is regular, and it reaches another asymptotically flat end as $z\to-\infty$.  The analytical extension of the metric components is shown in Fig. \ref{figmetric}. The causal structure is equivalent to Fig. \ref{CGHSboulwarenegativa}.

\begin{figure}[tbp]
\begin{center}
\includegraphics[angle=0,width=3.8in,clip]{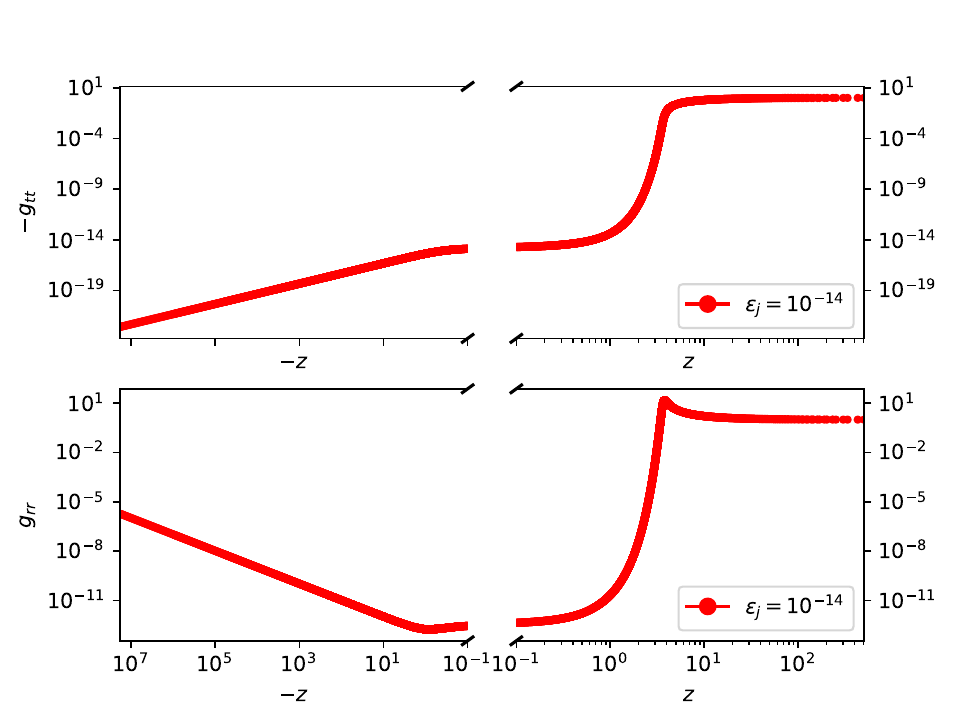}
\caption{Analytical extension of the two-dimensional spacetime metric components (\ref{2metric}) through negative values of $z$, obtained by solving (\ref{invfunc}) with asymptotically flat boundary conditions (\ref{z1}) and (\ref{z2}) for $C=-1$ and $a=2$.}  \label{figmetric}
\end{center}
\end{figure}

These results can be confirmed using standard techniques from the theory of dynamical systems \cite{stroglatz}. Namely, Eq. (\ref{invfunc}) can be displayed as a first-order Hamiltonian system of the form
\bea \label{hamsystem}
\partial_\rho z &=& T\, ,\\
\partial_{\rho} T&=&\frac{(z+T)(C\frac{\lambda}{r_0^2}+2z T)}{z^2-C\frac{\lambda}{r_0^2}}\ .
\eea
This formulation allows us to draw a stream plot for the ``velocity'' vector $(z_\rho,T_\rho)$ at each point $(z(\rho),T(\rho))$, called the phase space of the system, which is shown in Fig. \ref{fig21}. Given an initial data point $(z(0),T(0))$, the phase space tells us the qualitative behavior of the solution of (\ref{invfunc}) as $-\rho$ increases from $0$ to $+\infty$. With the asymptotically flat boundary conditions (\ref{z1}) and (\ref{z2}), $(z(0),T(0))$ is located in the upper right corner of  Fig. \ref{fig21} {\it for any value of $a>0$}. As $-\rho$ increases, the corresponding solution $(z(\rho),T(\rho))$ follows the flow indicated by the arrows. Figure \ref{fig21}  shows that it will approximate a curvature singularity (red curve) after crossing a point with $R=0$ (green curve), but will never intersect it. This is the maximum peak found in Fig. \ref{fig20} for the scalar curvature. After a while, the solution tends to $(z(\rho),T(\rho))\to (-\infty, +\infty)$ as $\rho\to -\infty$, for which $R\to 0$, i.e. the spacetime approaches an asymptotically flat end. Notice that this  behavior holds for any value of $a>0$, so we can extrapolate the qualitative behavior of the numerical results presented above for $a=2$ to astrophysically relevant values of the mass $M$. 

Finally, it is  interesting to note that, as mentioned above,  the numerical results do not exhibit the properties of a type I wormhole geometry, nor do they exhibit also the properties of a type II wormhole as defined in \cite{wormholetype}. We do not find a local minimum for the metric function $e^{2\rho}$. However, we leave open the interesting possibility of obtaining a type II wormholelike geometry  by considering  hybrid states  (i.e., some physical fields with positive central charge, but still an overall
negative central charge), as described for the RST model in \cite{PSSprl, PSS1, PSS2}.

\begin{figure}[htbp]
\begin{center}
\includegraphics[angle=0,width=3.8in,clip]{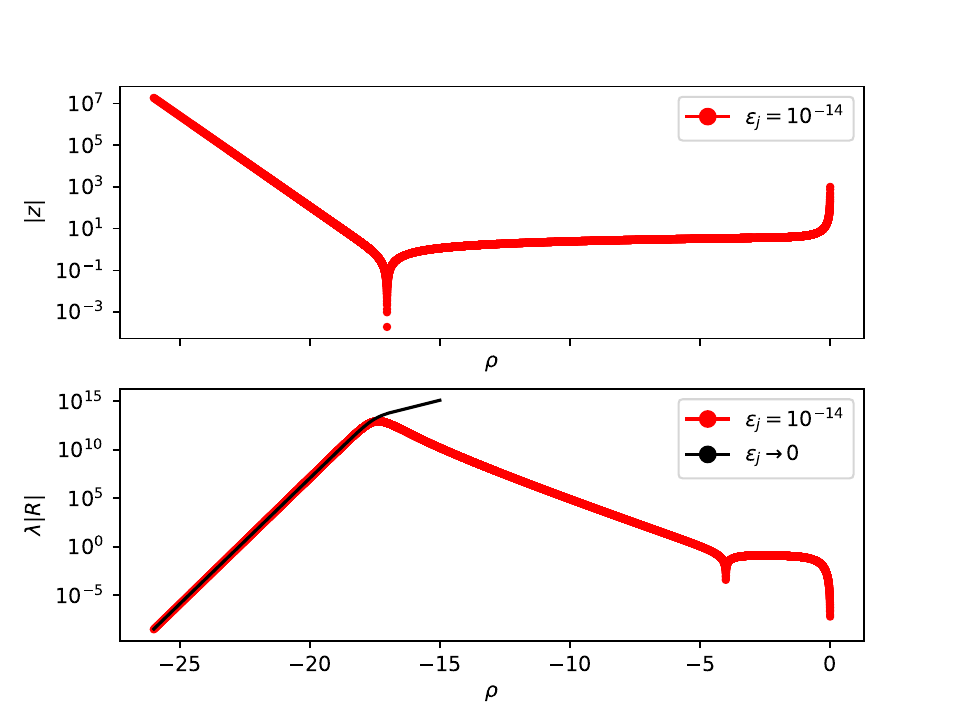}
\caption{Analytical extension of the two-dimensional spacetime metric (\ref{2metric}) through negative values of $z$, obtained by solving (\ref{invfunc}) with asymptotically flat boundary conditions (\ref{z1}) and (\ref{z2}) for $C=-1$ and $a=2$. The upper panel shows the inverse  function of $\rho=\rho(z)$, while the lower panel displays the radial profile of the scalar curvature. Notice how $|R|$ remains always bounded from above (i.e. no curvature singularity) and $R\to 0^\pm$ for both $z\to \pm \infty$ (i.e. asymptotic flatness at both ends). In the limit $\epsilon_j\to 0$ the curvature converges to $\lambda R(\rho) \to - 2 e ^{ |\tilde m| \rho + \tilde n}/ (z^2(\rho)+1) $ for $z(\rho) < 0$ and some real numbers $\tilde m$, $\tilde n$ (black line), while its error tends to zero.} \label{fig20}
\end{center}
\end{figure}

\begin{figure}[htbp]
\begin{center}
\includegraphics[angle=0,width=3.5in,clip]{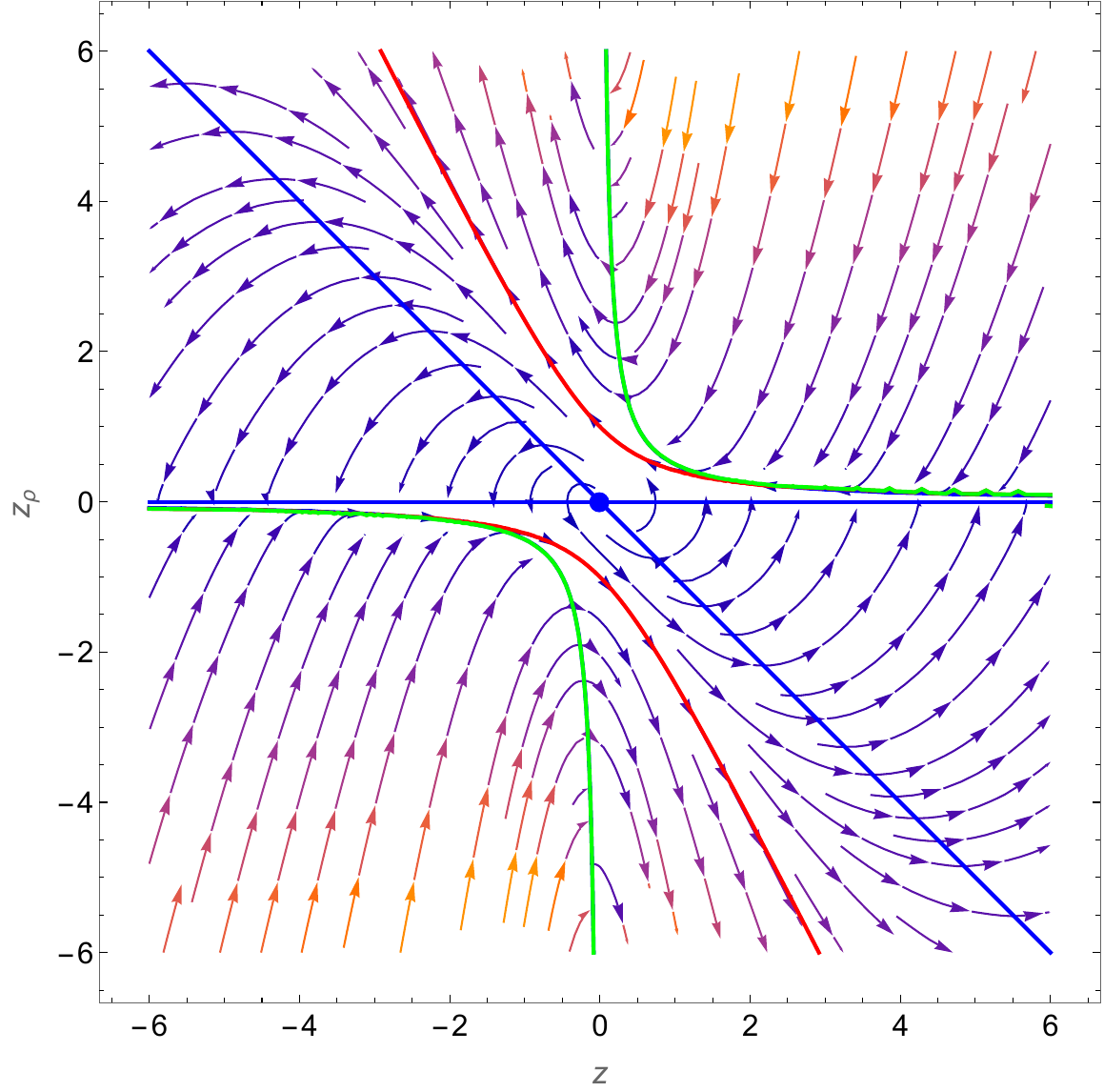}
\caption{Phase space of (\ref{invfunc}) with $C=-1$ and $\lambda=r_0^2$.  The green line  represents points with zero curvature, while the red line represents curvature singularities. The blue lines correspond to nullclines (i.e. points where either $z_\rho=0$ or $z_{\rho\rho}=0$), some of which agree with the green lines.  With  asymptotically flat boundary conditions (\ref{z1}) and (\ref{z2}), the system evolves from a point in the first quadrant {\it for any value of $a>0$}, then crosses a point with $R=0$ (green line),  approaches a curvature singularity (red line) and finally expands away from it. This is, precisely, the behavior found numerically for $a=2$ (see Fig. \ref{fig20}). Since the only fixed point of (\ref{invfunc}) is the origin, the system is expected to evolve to $z\to-\infty$ departing progressively from the red line, leading eventually to $R\to0$.} \label{fig21}
\end{center}
\end{figure}

\section{Conclusions and discussion}
\label{Conclusions}

The combination of the results of Secs. \ref{sectioncghs} and \ref{sectioneinstein} shows a simple pattern for black holes described within the framework of 2D dilaton gravity.  It was known that the backreaction of quantum fields in the Boulware state,  with positive central charge, removes the classical black hole horizon, but not the classical spacelike singularity, which persists in the semiclassical theory in the form of a null singularity. 
Here we have shown that, somewhat surprisingly, reversing the sign of the central charge of the two-dimensional conformal matter provides a mechanism for singularity resolution. This was recently found in \cite{PSSprl, PSS1, PSS2} for a semiclassical form of the  CGHS model, namely, the RST model. Because of the existence of a special global symmetry in the RST model, this question could be answered in a simple analytical way. Nevertheless, it could happen that the resolution of the singularity in this model may be an accidental feature of this special symmetry. Partially motivated by this,  in this paper  we have  focused  on a  more involved (but more conservative) model of 2D dilaton gravity, namely the  semiclassical, spherically symmetric reduced  theory of general relativity. Classically, this model  describes a Schwarzschild black hole, so it is more interesting from a physical point of view. Furthermore, the model has no additional global symmetry and the static semiclassical solution requires numerical analysis. Our results  indicate that the  classical singularity is resolved by quantum effects. 

This mechanism of singularity resolution is driven by a negative value of the central charge $C$ in the two-dimensional conformal anomaly,
\be \langle T_a^{\,\, a} \rangle = \hbar \frac{C }{24\pi} R \ , \ee
 which can be regarded as arising from nonphysical or exotic degrees of freedom, because unitarity   requires $C >0$ \cite{FQS84}.  This apparent drawback  can be mitigated by considering a somewhat analogous situation in four spacetime dimensions.  In this case, the trace anomaly is given by the general formula
\be \label{4danomaly} \langle  T^a_a \rangle = \hbar ( c \ W^{abcd}W_{abcd} -a  \ E )\ , \ee
where $W_{abcd}$ is the Weyl curvature tensor 
and $E$ is the Euler density.\footnote{We have ignored contributions of the form $\Box R$ as they are intrinsically ambiguous and can be shifted by local counterterms}
 The numerical coefficients $c$ and $a$ depend on the spin of the field, and they are positive for conventional spin-$0$, spin-$1/2$, and spin-$1$ matter fields \cite{birrell-davies, Duff}. However, in four dimensions it is still possible to have  quantum fields with negative $a$ and $c$ without breaking unitarity. 
To give a first example, it is well  established that the  gravitino field  contributes negatively to both the $a$ and $c$ coefficients \cite{CD1, CD2}.  A second example, less known, is 
 a conformally invariant scalar field $\xi$ of zero dimension, which also contributes negatively to both $a$ and $c$ \cite{Gusynin89}.
 This field is defined  by the action
 \be \label{Sxi} S= -\frac{1}{2} \int d^4x \sqrt{-g} \ \xi \triangle_4 \ \xi \ ,  \ee 
 where  $\triangle_4$ is the unique conformally invariant fourth order operator. 
The flat space version of this  theory has been studied in \cite{Bogolubov}. The theory has an underlying gauge invariance which prevents the existence of local degrees of freedom. The vacuum is the only physical state, and no physical excitations are allowed.  This strongly suggests that, in a classical black hole background, the vacuum state for the theory (\ref{Sxi}) should produce a quantum stress tensor $\langle T_{\mu\nu}\rangle$ which decays to zero at infinity. There is no room for nonzero values in the asymptotically flat regions. This means that this vacuum   behaves as  a stable Boulware-type quantum state. The extrapolation of the results of this work  suggests that the existence of  dimensionless scalar fields (see \cite{BT21, MVZ, BNNP, NS} for additional physical motivations) would necessarily lead  to the removal of the event horizon, and a completely new picture of the backreacted  spacetime geometry of evaporating black holes. These issues deserve to be explored in a future work.

\section*{Acknowledgments} \label{sec7}

A. D. R. acknowledges financial support via  ``{\it Atraccion de Talento Cesar Nombela}'', Grant No. 2023-T1/TEC-29023, funded by Comunidad de Madrid, Spain. F. J. M.-G. is supported by the Ministerio de Ciencia, Innovaci\'on y Universidades, Ph.D. fellowship, Grant No. FPU22/02528. This paper has been supported by Project No. PID2023-149560NB- C21 funded by MCIU /AEI/10.13039/501100011033 / FEDER, UE and by No. CEX2023-001292-S funded by MCIU/AEI.  The paper is based upon work from COST Action CaLISTA CA21109 supported by COST (European Cooperation in Science and Technology).

\end{document}